\documentclass[11pt]{article}

\usepackage{makecell,multirow,enumitem,subfig,bm,booktabs}
\usepackage[preprint]{jmlr2e}
\usepackage{amsmath,amssymb,mathtools}

\begin{document}

\title{lcpy: an open-source python package for parametric and dynamic Life Cycle Assessment and Life Cycle Costing}

\author{\name Spiros Gkousis$^*$ \email s.gkousis@imperial.ac.uk\\
        \addr Imperial College London,\\ Department of Civil and Environmental Engineering,\\ Exhibition Rd, South Kensington, London,\\ SW7 2AZ, United Kingdom\\
        \name Evina Katsou \email e.katsou@imperial.ac.uk\\
        \addr Imperial College London,\\ Department of Civil and Environmental Engineering,\\Exhibition Rd, South Kensington, London, SW7 2AZ,\\ United Kingdom\\
       }

\maketitle

\begin{abstract}%   <- trailing '%' for backward compatibility of .sty file
This article describes lcpy, an open-source python package that allows for advanced parametric Life Cycle Assessment (LCA) and Life Cycle Costing (LCC) analysis. The package is designed to allow the user to model a process with a flexible, modular design based on dictionaries and lists. The modeling can consider in-time variations, uncertainty, and allows for dynamic analysis, uncertainty assessment, as well as conventional static LCA and LCC. The package is compatible with optimization and uncertainty analysis libraries as well as python packages for prospective LCA. Its goal is to allow for easy implementation of dynamic LCA and LCC and for simple integration with tools for uncertainty assessment and optimization towards a more widened implementation of advanced enviro-economic analysis. The open-source code can be found at \url{https://github.com/spirdgk/lcpy}.
\end{abstract}


\section{Introduction}

Life Cycle Assessment (LCA) and Life Cycle Costing (LCC) analysis are widely adopted methods for assessing, respectively, the environmental and economic sustainability of products and systems. For a long time, LCA was limited by its static nature \citep{finnveden_2009} providing a snapshot of the environmental impacts without exploring the evolution of the impacts in time and how changes in the investigated system affects these. A few studies have explored dynamic approaches to LCA. For example, \cite{levasseur_2010} developed dynamic characterization factors for the global warming impact that account for the timing of the emissions in the impact calculation. \cite{collinge_2013} suggested adding a time dimension for solving the inventory equation to calculate a dynamic inventory and applied this approach for the LCA of buildings. \cite{bsp_2014} introduced the enhanced structural path analysis (ESPA) method for dynamic LCA using time distributions for modeling the environmental interventions of the investigated system. The resulting dynamic LCI could then be matched to dynamic characterization factors for the Life Cycle Impact Assessment (LCIA). Several other studies have applied dynamic LCA describing frameworks for its implementation, as well as suggesting dynamic characterization factors for numerous impact categories \citep{gkousis_2022, su_2017, lebailly_2014}. 

The development of the brightway software allowed for flexible LCA calculations in python \citep{Mutel_2017} and facilitated the development of open-source tools performing dynamic and prospective LCA. For example, the Premise software \citep{Sacchi_2022} can calculate, prospective Life Cycle Inventory (pLCI) databases that account for projections of integrated assessment models. These pLCI databases can then be used in brightway for prospective LCA and can be coupled with  dynamic LCI models. The Temporalis package \citep{Cardellini_2018} is a python tool for dynamic LCA based on brightway that proposes solving the dynamic inventory problem through graph traversal and convolution algorithms. In addition to Temporalis and Premise, more python tools are being developed to allow for advanced prospective and dynamic LCA \citep{Diepers_2025, Sacchi_2024}.
 
Despite the recent developments in dynamic and prospective LCA the operationalization of these methods is lacking, potentially because the developed tools require proficiency in the brightway software. At the same time, the developed tools have not focused on the system costs and the synergies between LCA and LCC towards a more holistic picture of the system sustainability. Lcpy suggests a simple approach to dynamic LCA and LCC which allows for parametric modeling, flexible time-step (e.g., years for prospective analysis or minutes for short-term analysis) and integration with uncertainty analysis and optimization libraries aiming to enable the widened adoption of advanced LCA and LCC of products and systems.

\section{Description}

\begin{figure}[t]
    \centering
    \includegraphics[width=0.8\textwidth]{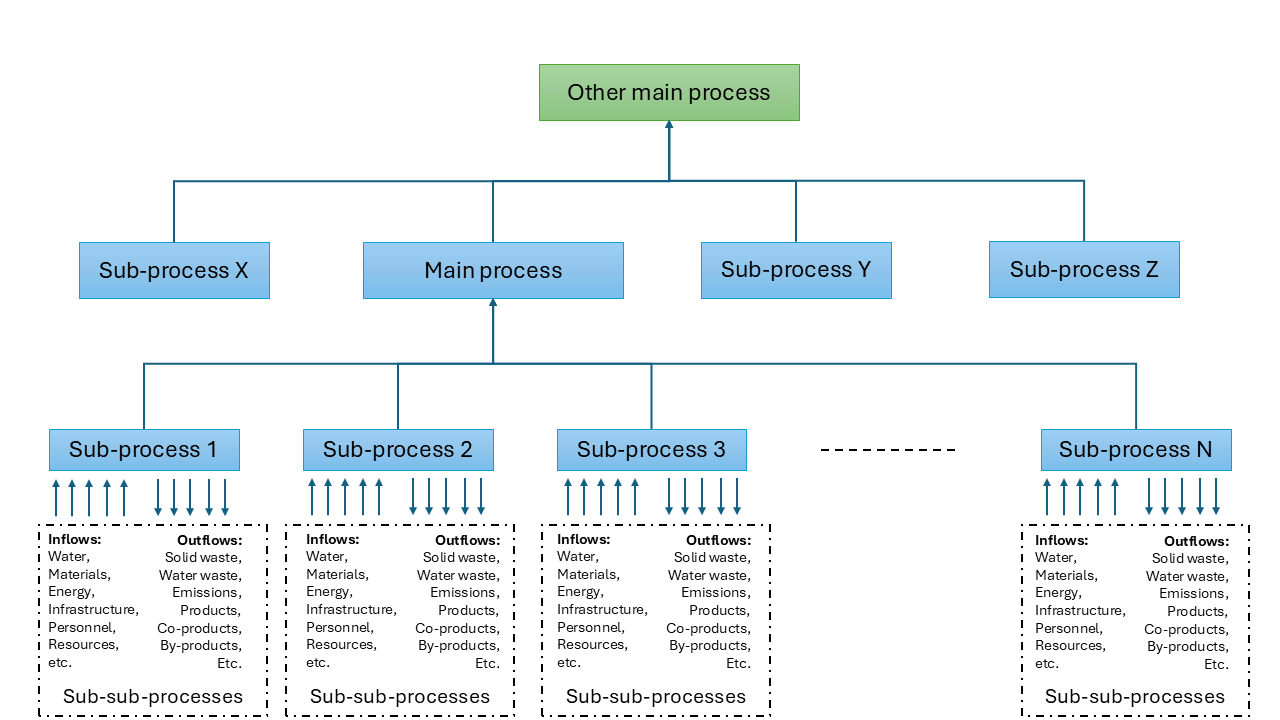}
    \caption{Approach to define the main process in lcpy.}
    \label{fig:fig_1}
\end{figure}

We consider that a main process/system is to be assessed. The main process is divided into several sub-processes, each consuming resources (inflows) and producing products, emissions, and waste (outflows) (Figure \ref{fig:fig_1}). Provided the exchange amounts, thus the amounts of each inflow and outflow for each sub-process and the amount of each sub-process in the main process, as well as the unit environmental impact and unit cost and revenue for each inflow and outflow, one can calculate the unit impact/cost of each sub-process and respectively of the main process. Equations \ref{eq:sp_unit_impact}-\ref{eq:mp_unit_cost} describe this procedure.  

\begin{align}
\text{SP}_{\text{unit-impact},i,j} &= \sum_{0}^{SSP} \left( \text{SSP}_{\text{unit-impact},i,j} \cdot \text{SSP}_{\text{exchange-amount},i,j} \right) \label{eq:sp_unit_impact} \\
\text{MP}_{\text{unit-impact},i,j} &= \sum_{0}^{SP} \left( \text{SP}_{\text{unit-impact},i,j} \cdot \text{SP}_{\text{exchange-amount},i,j} \right) \label{eq:mp_unit_impact} \\
\text{SP}_{\text{unit-cost},i,j}   &= \sum_{0}^{SSP} \left( \text{SSP}_{\text{unit-cost},i,j} \cdot \text{SSP}_{\text{exchange-amount},i,j} \right) \label{eq:sp_unit_cost} \\
\text{MP}_{\text{unit-cost},i,j}   &= \sum_{0}^{SP} \left( \text{SP}_{\text{unit-cost},i,j} \cdot \text{SP}_{\text{exchange-amount},i,j} \right) \label{eq:mp_unit_cost}
\end{align}
Where i,j are the number of scenarios and time steps considered, and MP, SP, and SSP are the main process, sub-process, and sub-sub-processes respectively.

Equations \ref{eq:sp_unit_impact}-\ref{eq:mp_unit_cost} are the core of lcpy \cite{gkousis_2025}. The package requires defining the names of the main process, sub-process, and sub-sub-processes (inflows and outflows) in python dictionaries with predefined structure. In addition, the exchange amounts for each sub-sub-process and sub-process need to be defined and passed into python lists. Depending on the type of analysis to be conducted, the exchange amounts can be inputted as simple values (simple LCA/LCC), or 2-D numpy \citep{harris2020array} arrays (Monte Carlo or dynamic LCA/LCC). Different calculator classes are provided to handle the 2-D or simple calculations, with the former being handled by cython functions to enable fast calculations. Next, the unit impacts and costs/revenues need to be derived. The latter can be derived as the market prices for each inflow and outflow and for the final products produced. The unit impacts for each inflow and outflow will vary depending on the impact category assessed (e.g., global warming impact, acidification) and the impact assessment method used. These can be calculated using a dedicated calculator class of lcpy that utilizes brightway2.0 \citep{Mutel_2017}. The user needs to create a brightway2.0 environment outside of the lcpy environment and create a project. Then, the key ids for the brightway2.0 processes that represent the inflows and outflows can be passed, as keys, to the python dictionaries that model each sub-process. The calculator class can then be used to derive the unit impact or inventory for each inflow and outflow. Alternatively, this information can be inputted by the user externally. Finally, a number of additional parameters need to be defined such as the number of scenarios and time-steps to be considered, impact assessment methods to be used, and discount rate for the economic analysis. Lcpy then provides functions to perform:

\begin{enumerate}
    \item Simple LCA and LCC/TEA
    \item Monte Carlo LCA and LCC
    \item Dynamic LCA and LCC
\end{enumerate}

Detailed examples are provided in the package repository \url{https://github.com/spirdgk/lcpy} and zenodo archive \citep{gkousis_2025} and detailed instructions and explanations for each function can be found in the documentation provided in the repository. Dynamic impact assessment can be performed using characterization factors with a predefined time horizon or with an annual time step as in the framework of \citep{levasseur_2010}. The dynamic characterization factors are inserted to lcpy as excel files. The construction of the excel files containing the dynamic characterization factors is explained in the documentation of the package \citep{gkousis_2025}. The calculation framework followed for the dynamic impact assessment is that described by \cite{gkousis_2022}, though the impact of substances characterized with static factors is assumed to be caused at the year of emission. Notably, in addition to simple LCC and cost analysis, lcpy provides functions to calculate the net present value, minimum selling price and the levelized cost of electricity.

In summary, lcpy consists of \textbf{(1)} a module which includes classes that perform the LCA and LCC calculations and hold the results, with different classes being available for dynamic and static calculations, \textbf{(2)} a module that provides functions for visualization and storing the results, and \textbf{(3)} a module that provides functions for the calculations used by the classes of the first module. In addition, a brightway 2.0 environment can be installed at a separate location and the path along with the name of the brightway project and database used for the background modeling of inflows and outflows need to be passed to the functions of lcpy through a structured dictionary.

\section{Impact and future steps}

The aim of lcpy is to enable low-cost dynamic LCA and LCC integrated with advanced computational methods such as optimization and global sensitivity analysis. It does not aim to replace existing packages for dynamic and parameterized LCA but to be used in combination with them. For example, dynamic LCIs can be built in other packages and the result can be used in lcpy to leverage its computational functions. It is to be used for building parametric models for process simulations and calculate the environmental and economic impact under various scenarios and evolving conditions. Also, it provides a simple way of performing optimization and uncertainty analysis (e.g., global sensitivity analysis) of the LCA and LCC results of generic systems and product by being compatible with optimization libraries such as Pymoo \citep{pymoo} and uncertainty analysis libraries such as SAlib \citep{salib} (see relevant example in \url{https://github.com/spirdgk/lcpy}). Next steps to its development are the incorporation of publicly available LCI datasets, and the development of excel files with pre-calculated dynamic characterization factors to allow for easy dynamic impact assessment. In addition, machine learning and artificial intelligence algorithms can be incorporated to fill data gaps and facilitate the matching of the sub-sub-processes to the keys ids of brightway processes, reducing thus the cost of performing the analysis.

\bibliography{ref}

\end{document}